\documentclass[12pt,preprint]{aastex}

\begin{document}

\title{XTE J1550-564: INTEGRAL Observations of a Failed Outburst}

\author{S. J. Sturner and C. R. Shrader}

\affil{Laboratory for High Energy Astrophysics, Code 661, 
NASA,Goddard Space Flight Center, Greenbelt, MD
20771}
\and
\affil{Universities Space Research Association, 10211 Wincopin Circle, 
Suite 620, Columbia, MD 21044}

\begin{abstract}
The well known black-hole X-ray binary transient XTE J1550-564 underwent an
outburst during the spring of 2003 which was substantially underluminous in comparison 
to previous periods of peak activity in that source. In addition, our analysis 
shows that it apparently remained in the hard spectral state over the duration
of that outburst. This is again in sharp contrast to major outbursts of that 
source in 1998/1999 during which it exhibited an irregular light curve,
multiple state changes and collimated outflows. This leads us to classify it
as a ``failed outburst." We present the results of our
study of the spring 2003 event including light curves based on 
observations from both {\it INTEGRAL} and {\it RXTE}. In addition, we studied the 
evolution of the high-energy 3-300 keV continuum spectrum using data obtained with three main
instruments on {\it INTEGRAL}.
These spectra are consistent with typical low-hard-state thermal Comptonization emission. 
We also consider the 2003 event in the context of a multi-source,
multi-event period-peak luminosity  diagram in which it is a clear outlyer.
We then consider the possibility that the 2003 event was due to a 
discrete accretion event rather than a limit-cycle instablility.  In 
that context, we apply model fitting to derive the timescale for viscous propagation in the
disk, and infer some physical characteristics.
\end{abstract}


\section{INTRODUCTION}

Black hole X-ray transients, a subset of the class of binaries known as 
X-ray novae or soft X-ray transients \citep{csl97, mr04}, are low-mass X-ray 
binaries (LMXBs) in which long periods of quiescence, often
decades, are interupted by dramatic X-ray and 
optical/UV outbursts, frequently accompanied by radio emission which is
sometimes associated with collimated outflows.
The X-ray binary XTE J1550-564 is an example of this class of objects.
It was discovered in 1998 \citep{smit98}, at which time it underwent a period of major outburst, surging to 
perhaps its Eddington luminosity assuming the distance and 
mass estimates of \citet{oros02}, 5.3 kpc and 9.4 $M_{\odot}$, respectively.  
This initial event, which exhibited a variety of complex behavior including an
irregular light curve, spectral state transitions, and quasi-periodic
oscillations, lasted for  $\sim$200 days \citep[e.g.][]{sobc00, homa01, remi02}. 
Radio monitoring during the X-Ray outburst recorded a radio flare, and subsequent VLBI radio 
observations soon thereafter showed evidence for a superluminal jet ejection event \citep{hann01}.
Subsequently, the jet has been detected in X-rays using {\it Chandra} and has 
been shown to be decelerating \citep{corb02,toms03,kaar03}.
Given these discoveries,  XTE J1550-564 is a strong Galactic black hole candidate and it has been classified 
as a microquasar.

Another major,  although less luminous than the first, outburst occured about 1.5 years 
later.   Since mid-2000, the source has remained active, 
exhibiting spoaradic, low-amplitude flaring episodes.  It has been 
suggested that these occur at nominal 1-year intervals, but this 
pattern is not well quantified. A long-term light curve illustrating 
these basic features is presented in Figure 1.

Early in 2003, during a Galactic plane monitoring scan, the 
INTErnational Gamma-Ray Astrophysical Laboratory ({\it INTEGRAL}) 
\citep[see][for 
a description of INTEGRAL and the Galactic plane monitoring program] {wink03}, detected 
the onset of outburst activity in XTE J1550-564 in the hard ($>$20 keV) 
X-ray band \citep{dubu03, arefiev04}. Subsequently, a series of 
ToO observations were carried out.
 The source was also observed with the Rossi 
X-Ray Timing Explorer ({\it RXTE}), both with
pointed observations as well as essentially continuous monitoring with the All-Sky
Monitor (ASM). In this paper we present our analysis and interpretaion 
of these data. 

We briefly discuss the instruments onboard {\it INTEGRAL} in \S 2.  
Our analysis of the {\it INTEGRAL} and {\it RXTE} observations is presented in 
\S 3 . This includes light curve analysis as well as our spectral model fitting and interpretation. 
 In \S 4 we discuss the 
low-amplitude hard-spectral state outburst of XTE J1550-564 in the 
context of models for X-ray nova outbursts, and spectral state 
transitions. Some conclusions are drawn in \S 5.

\section{INSTRUMENTATION - INTEGRAL}

{\it INTEGRAL} was launched into a 3-day elliptical orbit on
October 17, 2002.  It carries four instruments, three of which have been 
used for our analyses: the imager IBIS \citep{uber03}, the spectrometer SPI \citep{vedr03}, 
and the Joint European X-Ray Monitor, JEM-X \citep{lund03}. For a complete 
description of the {\it INTEGRAL}
spacecraft and mission refer to \citet{wink03}.

IBIS is a coded mask instrument which has a wide field of view
(FOV) of 29$^{\circ}$ $\times$ 29$^{\circ}$ (9$^{\circ}$ $\times$  9$^{\circ}$ fully coded) 
with a point spread function
(PSF) of 12$'$ (FWHM) and is sensitive over the energy range
15 keV to 10 MeV. There are  two detector layers: ISGRI, an upper CdTe layer 
with peak sensitivity between 15 and
200 keV, and PICsIT, a bottom CsI layer, with a peak
sensitivity above 200 keV.  In this paper we
have used only ISGRI data. 

The Spectrometer on Integral, SPI, covers the 20 keV - 8 MeV energy
range with an energy dependent resolution of  2-8 keV (FWHM).
It consists of an array of 19 hexagonal high-purity Germanium
detectors. A hexagonal coded aperture mask located
1.7 m above the detection plane images large regions of the sky
(fully coded field of view  16$^{\circ}$) with an angular
resolution of $\sim$2.5$^{\circ}$.

The Joint European X-ray Monitor telescope, JEM-X, detection plane
consists of two identical high-pressure imaging microstrip gas
chambers. Currently, only one telescope/detector 
is in operation at a time. Each detector views the sky through a 
coded aperture mask. During the observations analyzed in this work, only
the JEM-X 2 telescope was operational.

\section{DATA ANALYSIS AND RESULTS}

We have analyzed the SPI, IBIS/ISGRI and JEM-X data
collected between January 29, 2003 and April 12, 2003 in a series of dither pointings
or ``Science Windows" (SCWs) lasting about 30-40 minutes each.
These include Core Program data from the Galactic Plane Scans (GPS) and
 the Galactic Center Deep Exposure (GCDE) as well as data from a
series of Target of Opportunity (ToO) observations activated as part of
the General Program  \citep[see][]{wink03}.  Data reduction
was performed using the standard OSA 4.1 analysis software package
available from the {\it INTEGRAL} Science Data Centre.  The data were
downloaded from the HEASARC mirror to the {\it INTEGRAL} Public Data Archive.
Only SCWs with a pointing direction within a specified angular radius were accepted.
This radius varied for each instrument.  We chose to use a 12$^{\circ}$ radius for SPI,
a 10$^{\circ}$ radius for IBIS/ISGRI, and a 2$^{\circ}$ radius for JEM-X.  These large
angular cuts were only relevant for SPI and ISGRI data prior to the outburst.  During the ToO
pointed observations the angular distance off-axis was never greater than $6^{\circ}$.
In addition  to the {\it INTEGRAL} data, we obtained the relevant
{\it RXTE}/ASM data from the ASM light curve archive.

\subsection{Imaging and Light Curves}

The long term X-ray light curve for XTE J1550-564 is shown in Figure 1.
The {\it RXTE}/ASM daily summations depicted there show the highly variable
nature of this source.  The major outbursts in 1998-1999 \citep{smit98} and 2000 \citep{rct03} are
very prominent.  The current outburst, indicated by the
vertical dashed line, is quite modest in comparison.
The January - April 2003 light curves for XTE J1550-564 as seen by IBIS/ISGRI and {\it RXTE}/ASM
are depicted in Figure 2.   In panels 2b - 2d we show the ISGRI light curves for 
the 15-40 keV, 40-100 keV, and 100-200 keV bands,
respectively. Each data point represents the average count rate for
an individual SCW, $\sim$ 2000 seconds.   In panel 2a we show the {\it RXTE}/ASM 1.5 - 12 keV 
light curve.  Each
data point there is for a daily average.

The {\it RXTE}/ASM light curve shows that this outburst began
on or about March 21, 2003 (MJD 52719) and lasted for roughly 45 days.  There is a broad peak lasting
$\sim$ 20 days with a peak count rate of 4.86 cts s$^{-1}$ or $\sim$64 mCrab.   
The IBIS/ISGRI light curve is less continuously sampled than the {\it RXTE}/ASM one.  The
outburst was first detected with ISGRI during {\it INTEGRAL} revolution 54 on March 24, 2003
during a GPS observation.  An {\it INTEGRAL} ToO was triggered and pointed observations commenced during
revolution 55 on March 27, 2003 and continued during revolutions 57 and 58, and ended during 
revolution 60 on April 12, 2003.   Hence there is no ISGRI data during the initial rise phase nor any
during the decay.  The shapes of the light curves in the three ISGRI energy bands are very similar
but there is a suggestion that the rise in the 100-200 keV band was slightly faster than the lower energy
bands.  The peak count rates in the 15-40 keV, 40-100 keV, and 100-200 keV energy bands are
37.6 cts s$^{-1}$, 35.2 cts s$^{-1}$, and 9.11 cts s$^{-1}$, respectively.  
In Figure 3 we show the IBIS/ISGRI mosaic image for revolution 57
which clearly shows XTE J1550-564 during this time period.

The major outbursts of 1998-1999 and 2000 showed transitions from a power law dominated low hard state to
a high soft state that had a strong thermal component.   We looked for similar transitions in the 2003 outburst
by examining hardness ratios.
Using {\it RXTE}/ASM data, we calculated hardness ratios for the large September 1998 - January 1999 outburst 
as well as the March-April 2003 outburst.  In this paper we define the hardness ratio when 
comparing two energy bands as (H-S)/(H+S) where H is the count rate in the higher energy band and
S is the rate in the lower energy band.   In Figure 4 we compare the hardness ratios for the two different outbursts
using the 1.5-3 keV (A) and 5-12 keV (C) energy bands of the {\it RXTE}/ASM.  For the March-April 2003 outburst
the hardness ratio is approximately constant with a value of $\sim$0.3.  In contrast, the hardness ratio during the
large September 1998-January 1999 burst varied significantly over its duration.  During the initial flare 
the hardness ratio was similar to the 2003 burst with values near 0.3.  The hardness ratio quickly
decreased to an almost constant value near -0.1.  During the second peak in January 1999, the hardness ratio
was even smaller with values near -0.3.  This in agreement with \citet{sobc00} who noted that the first peak of the outburst
was generally dominated by power-law emission while the second peak was dominated by softer thermal disk emission.
Thus we find that not only is the March 2003 event of signifcantly lower peak
luminosity, it is also appears to never leave the canonical low hard state.  We will examine this further in the 
next section when we discuss our spectroscopy results.

\subsection{Spectral Analysis }

We performed spectral analysis of the JEM-X,
IBIS/ISGRI, and SPI data during the March-April 2003 outburst using OSA 4.1 software.  
For JEM-X and ISGRI, individual spectra are produced
for each SCW.  For strong sources such as XTE J1550-564 during this outburst, these
individual SCW spectra can be easily combined to improve there statistical significance and for direct
comparison with SPI spectra.
We used the OSA utility program {\it spe\_pick} to combine the JEM-X spectra by revolution.  The IBIS/ISGRI
spectra were similarly combined using the FTOOL {\it mathpha}.
SPI is a coded mask telescope like IBIS and JEM-X.  But unlike these other instruments, 
it is not possible to extract spectra for individual SPI pointings due to its relatively small number
detectors and mask elements.  Thus for SPI one can only extract spectra for a series of pointings.  In this case we chose to
 extract one spectrum for each of revolutions 55, 57, and
60  using all the relevant SCWs  in a given revolution.   We then performed a simultaneous fit of the SPI, IBIS/ISGRI, and
JEM-X data, for each of the 3 revolutions, in XSPEC v.11.3.1.
 We explored fits using the COMPST \citep{st80} and COMPTT \citep{t94} thermal Comptonization models.  We also 
 explored the need for an additional spectral component due to reflection of the hard Comptonized spectrum off of the
 relatively cool portion of the accretion disk using the PEXRAV model \citep{mz95, gcr99}.
During the fitting procedure the model normalization was allowed to vary from instrument to instrument
and a systematic 5\% error was added to the IBIS/ISGRI data in order to 
achieve reasonable $\chi^2$ values \citep[see e.g.][]{gold03}. 
Since the data being fit were restricted to energies greater than 3 keV, the hydrogen column density is not well constrained
by the data.  Therefore in the fitting process we fixed the column density at $N_{\rm H} = 9 \times 10^{21}\ {\rm cm^{-2}}$.  
This value for $N_{\rm H}$ was derived using the FTOOL NH which uses the results of 
\citet{dl90} and is fully consistent with the values derived by 
\citet{kaar03} using {\it Chandra} observations of the jets of XTE J1550-564.  The PEXRAV model contains the 
system inclination angle as parameter.  We set this parameter to $72.6^{\circ}$ as determined by \citet{oros02}.

We found that the COMPTT model provided a significantly better fit to the data for each 
revolution than the COMPST with $\Delta \chi^2 \gtrsim 40$.  The COMPTT tends to produce lower residuals above $\sim$100 keV.   In terms of quality of fit, the results using the PEXRAV model fell in between the COMPTT, $\Delta \chi^2 \lesssim 25$, and COMPST  results.  The complete results of our fitting procedures are given in Tables 1 \& 2. 
Given these results we cannot make any determination about the need for a reflection component when fitting
the spectrum of XTE J1550-564 during this outburst.  We also note that there is no compelling evidence for 
flourescent iron-line emission within the S/N limits of our data.
  There is a significant yet consistent normalization difference
between JEMX, IBIS/ISGRI and SPI with ratios of approximately 0.9:1.0:1.6.  
These cross calibration issues have been previously noted, see e.g. \citet{cour03}.

\citet{arefiev04} have published results from the early rise phase of this outburst using {\it INTEGRAL} Core Program and
{\it RXTE}/PCA data.
We find that our best fit plasma temperatures and optical depths are similar to theirs while our 3-200 keV
model flux for revolution 57 ($9.07 \times 10^{-9}$ ergs cm$^{-2}$ s$^{-1}$ using the ISGRI normalization) 
is roughly 2.4 times higher than theirs.  
This is expected since the vast majority of their data was from before MJD 52725 when the IBIS/ISGRI count rate
was a factor of 2-3 below the peak (see Figure 2).  We also note that the reflection scaling factors we obtain when 
fitting with the PEXRAV model are consistent with that found by \citet{arefiev04}, $0.25\pm 0.13$, although the errors in both instances  are rather large (see Table 1).  There is also some evidence for spectral evolution to flatter power laws, lower cut-off energies, and a smaller reflection contribution with time but the errors on these parameters are significant.

In Figures 5 \& 6 we show the spectral fits to the data from revolution 57 using the COMPTT model. 
In Figure 7 we show the 90\% confidence contours
for the $kT$ and $\tau_{\rm p}$ COMPTT model parameters for each revolution.  There is a strong suggestion of
spectral evolution to lower $kT$ and higher $\tau_{\rm p}$ with time.
No Fe line structure is evident at the resolution and sensitiviy of JEM-X. 

Thus the spectra from each of the three revolutions are well fit by the COMPTT thermal Comptonization 
model which produces essentially a 
cutoff power-law spectrum.  We again find no indication of a transition from the low hard 
state to a disk dominated high soft 
or intermediate state as was seen in the major outbursts in 1998-1999 \citep{sobc00} and 2000 \citep{rct03}.
We also note the absense of any statistically significant excess relative to the model at the highest energies,
as has been noted for other sources \citep[e.g.][]{grove98}. Such excesses may have to do with the 
radial density profiles of the Comptonizing plasma, and the inadequacy of the analytcal 
forms of the models to take such effects into account
upwards of several hundred keV \citep{demos97}.  Alternatively, such excess continuum 
could be associated with an additional emission component resulting from bulk motion Comptonization (Titarchuk 2004,
private communication).

Finally, in addition to the static thermal Comptonization models, we considered Comptonization by an
expanding plasma ambient to the central compact object. It was recently shown that the presence of 
such an outflow is likely to produce a net down-scattering of the emerging photon field -- thus, a
hardening of the spectrum in the $\sim20-100$ keV region (Titarchuk \& Shrader 2004). 
The observable effect is similar to what would be expected from various Compton reflection 
scenarios. As noted however, there does not appear to be a statistically significant continuum
residual at those energies in our static thermal Comptonization model fits.
Since plasma ejection as evidenced from radio emission is believed to accompany the high-soft to low-hard
state transition, one might conjecture that the hard-state turn on in these types of flares could also
involve outflowing plamsa. We found that the model described by Titarchuk \& Shrader (2004) led in some
cases to a slightly improved fit; for revolution 57 for example, we obtained 271/205 compared to 
287.7/205). We thus suggest that our results are consistent with scenarios involving plamsa ejection.
However,  the statistical improvement to the fit is marginal, thus, particularly in the absense
of radio emission, we are reluctant to draw strong conclusions.

\section{DISCUSSION}

\citet{rct03} showed that during the April 2000 outburst of XTE J1550-564 there was spectral 
evolution from the low hard state to a softer 
intermediate state as the luminosity increases and then back to the low hard state as the 
source luminosity declines.  The transitions
from one state to another were rather abrupt and were described as a phase transition with hysteresis, i.e. the
hard-to-soft transition occurs at at a higher flux than the soft-to-hard transition.   This hysteresis effect has been 
previously noted, e.g. \citet{miya95}.
The initial transition from the low hard state to the intermediate state was shown to occur when the 
 integrated 2 - 200 keV flux reached $\sim 2.3 \times 10^{-8}$ ergs cm$^{-2}$ s$^{-1}$. 
  The transition from the intermediate state back to the low hard state for the 2000 outburst 
 occurred a flux of $\sim 1.3 \times 10^{-8}$ ergs cm$^{-2}$ s$^{-1}$.
 Comparing the 2000 and 2003 outbursts,  the  2-200 keV integrated flux for the March 2003 outburst never exceeded
 $\sim 10^{-8}$ ergs cm$^{-2}$ s$^{-1}$, thus never reaching the flux where \citet{rct03} saw a transition to the
 intermediate state. 

 Assuming a distance to XTE J1550-564 of
 5.3 kpc and a black hole mass of 9.4 $M_{\odot}$ \citep{oros02}, the corresponding 2 - 200 keV luminosities at which 
 the hard-to-soft (H-S) and soft-to-hard (S-H) transitions occurred during the the 2000 outburst are 
 $L_{\rm H-S,2000} = 7.7 \times 10^{37}$ ergs s$^{-1}$ $= 0.063 L_{\rm EDD}$ and 
 $L_{\rm S-H,2000} = 4.4 \times 10^{37}$ ergs s$^{-1}$ $= 0.036 L_{\rm EDD}$.
 During the 1998-1999 outburst, the transition from the high soft state to the low hard state occurred at
 a luminosity $L_{\rm S-H,1998} = 0.034 L_{\rm EDD}$ \citep{macc03,sobc00}.
 Thus the luminosity at which the soft-to-hard transition occurs is consistent between the two major outbursts.
 They are also consistent with theoretical predictions of state transitions 
 at luminosities near $0.02 - 0.05 L_{\rm EDD}$ if one accounts for the hysteresis effect \citep{meye04,mm00}.
 For the 2003 outburst, we find that the 2 - 200 keV flux  during revolution 57 was $9.33 \times 10^{-9}$ ergs cm$^{-2}$ s$^{-1}$
 if we use the ISGRI normalization for the COMPTT model.  This corresponds to a 2 - 200 keV luminosity of 
 $3.1 \times 10^{37}$ ergs s$^{-1}$ $= 0.025 L_{\rm EDD}$ which is toward the lower bound of the theoretical range 
 of luminosities where a transition could occur.  Thus the 2003 outburst of XTE J1550-564 
 could be classified as a failed major burst in which the  luminosity never reached the point of spectral transition.

The fact that XTE J1550-564 remained in it low-hard spectral state during 
the 2003 outburst (and possibly in other minor outbursts over the last 
several years) poses some interesting challenges to the 
disk-thermal-instability model for outbursts. This is the case both 
because of the relatively long binary period, 1.54 days \citep{oros02} with the implied large separation
(and accretion disk size), combined with the  rather high frequency of the 
outbursts. 

In one commonly accepted view of black hole X-ray transients, the low-mass companion star
undergoes roche-lobe overflow, but at a sufficiently low rate that the
gas is accumulated in the (cool) disk until a critical level is reached, at which point the 
outburst occurs \citep[e.g.][]{laso01, meye04}. Just what this critical 
level of accumulated mass is unclear, but it is certain to depend on the 
size of the disk, and thus the binary separation. This behavior is
analogous to the outburst cycles seen in dwarf novae, however the
recurrence timescales for X-ray transients are generally longer and less regular.  In between
outbursts, the accretion rate is low, and it is conjectured that an
advection-dominated region fills the inner disk. The spectrum is hard
(photon indices $\Gamma\sim 1.5-1.9$) and consistent with models
of thermal Comptonization.  Most often, in outburst the spectrum
changes to a $\sim1$ keV thermal spectrum, often interpreted as the radiation
from a geometrically thin, optically thick disk, superposed on a softer 
power law (photon indices $\Gamma\sim 2-3$) extending to 10's or 100's of
keV. Such spectral state transitions are also well known in persistent X-ray binary sources.

It has been suggested \citep[e.g.][]{nowa95, meye04}
that whether or not a spectral-state transition occurs
depends on the value of $\dot{m}$.  In the ADAF model, the mass inflow 
rate must be sufficient to overcome the transition to the two-temperature 
(electron-ion) plasma, so that it can extend close to the
innermost stable orbit \citep{mm00}.  The details of this process
may be crucial to a complete understanding of spectral-state
transitions.  It may be the case that for low $\dot{m}$ regimes, sources
remain in the low hard state.   \citet{meye04} have
suggested that short-period systems (short in this context being $P
\sim$5 hours or less), by virtue of their smaller binary separations and
correspondingly smaller accretion disks,  would naturally have outbursts
of lower luminosity.   Additionally, the efficiency with which the ADAF
corona is produced may depend on the inner disk radius, and may be
bounded by some maximum value.

Historically, it is notable that a few of the transient LMXBs  remain
in the low hard state during outburst, e.g. GRO J0422+32, GRO J1719-24 and
XTE J1118+480 (e.g. Brocksopp 2001). But in two of these cases at least,
the orbital periods are small, 4.1 hours for 
XTE J1118+480 and 5.1 hours for GRO J0422+32 \citep{mr04}.  Thus implying
small accretion disks and accumulated reservoirs of matter.
Given XTE J1550-564's  relatively long orbital period, the disk is large among LMXBs; 
$R_d\sim3.5-4.0\times10^{11}$ cm. 
Thus the mass accumulated in the accretion disk between outbursts should be
correspondingly large. How then are these small outbursts produced? 
One possible explanation is that
discrete accretion events can somehow occur in which only 
part of the inner disk is disrupted.

In view of the substantial differences between the major 
outbursts and the subsequent minor flaring events -- in 
terms of the peak luminosities, the light curve shape and
duration, as well as the hardness ratio and the recurrence frequency 
of similar events -- one might speculate that some mechanism other than the 
thermal limit cycle could be occurring. 

For the spring 2003 event, we can estimate the total 
mass transfer based on our spectral analysis, and the
a simplistic assumption that the spectral energy distribution remains nearly constant
in shape, with the amplitude tracing the light curve envelope (see Figure 8).
This leads to a fluence of about $7\times10^{43}$ ergs. For an efficiency 
of 10\%, this corresponds to about $3.8\times10^{-10}$ solar masses of
material transferred. From Figure 1, its is evident that the events of this 
magnitude are ocurring on approximately annual time scales. The major outbursts of
1999 and 2000 execeed the 2003 event by several orders of magnitude
in fluence and are more closely spaced. Thus, the disk-thermal instability
models must be able to reproduce a wide variety of phenomena. We suggest that
there may be more than one mechanism functioning in XTE J1550-56. 
For example, a small, discrete
accretion event, ie. a discontinuity in the mass transfer rate which would lead
in a natural way to a ``fast-rise, exponential decay" (FRED) 
light curve. Persuing this hypothesis, 
we applied the Diffusive-Dissipation Propagation model \citep{wood01} to 
the {\it RXTE}/ASM light curve data.
This model predicts a FRED-type light curve. The resulting model fit is shown in 
Figure 8 where the inferred 
paramater of the fit, the diffusion timescale $t_0$,  was about 71 days. 
The diffusion timescale is defined as $t_0 = 16R^2/3\nu(r)$ and is roughly 
two times the characteristic light-curve decay timescale. Assuming the 
disk is about half of the roche radius \citep[e.g.][]{jpa96}, and 
using the known orbital parameters, we estimate that the accretion 
disk outer radius is $\sim2\times10^{11}$ cm. This leads to a viscosity 
of $\sim 10^{16}$ cm$^2$/s.

Finally, we note one additional possibility. It was recently pointed out by
\citet{nm04}, that XTE J1550-564 is among a subset of transient 
BH X-ray binaries, known to 
have high inclination angles ($i\gtrsim70^{\circ}$). Each of these objects 
have exhibited irregular outburst light curves following there initial turn-on.
The two early outburst peaks, with an intermittent valley of near-zero intensity
seen in Figure 1 are in that sense qualitatively similar to the outburst light curves 
of GRO J1655-40 and XTE J1118+48 for example. As \citet{nm04} note, 
SAX J1819.3-2525 which has the highest known inclination ($i\simeq75^{\circ}$)
underwent even more extreme erratic behavior, with separate outburst peaks spanning 
more than two orders of magnitude. This leads to the speculation that this
irregular behavior may be a result of variable column densities along our
line of sight, rather than to variations in the underlying outburst physics. 
For example, a flared outer disk of varying geometry, or with low-frequency
precessional modulation could lead to such scenarios.
It could perhaps then be the case that the hard, low-amplitude outbursts 
occur at times when the central X-ray source is highly obscured. What we are 
seeing in that case could be the Compton-scattered emission from a spatially
extended coronal region. However, the central thermal source in such scenarios
would have to be obscured by columns of $\sim10^{24}\ {\rm cm^{-2}}$ or else 
its presence would be revealed in the broad-band X-ray spectral analysis.

\section {Conclusions}

We have examined the flux history and spectral characteristics of 
the March/April 2003 outburst of XTE J1550-564, with emphasis on its
the high-energy observational coverage provided by {\it INTEGRAL}. Our major 
conclusions can be summarized as follows:

\noindent (i) It was an underluminous event, with a fluence more then two orders of magnitude
lower than the major outbursts of that source in 1999 and 2000. There appear
to be a number of such events in the long-term {\it RXTE}/ASM light curve data,
which ocurr with a frequency of 300-500 days. A recurrence rate of this scale,
combined with the extreme range in peak luminosity, would seem to pose
significant constraints on the standard thermal-disk instability model
of X-ray nova outbursts.

\noindent (ii) The source apparently remained in the low-hard spectral state
throughout the duration of the outburst. There was relatively little spectral
evolution, although there is evidence for a moderate softening of the
spectrum towards the end of our coverage. 

\noindent (iii) The spectral energy distribution is well approximated by a thermal 
Comptonization form. The inferred parameters, $\tau\simeq1.5$, $kT\simeq50$ keV, 
are similar to various other low-hard-state X-ray binaries. An alternative 
Comptonization model, which involves scattering in a divergent outflow, was fit to the data. While 
not conclusive, results were of comparable statistical quality to those obtained
for static Comptonization models, consistent at least with the presence of outflowing plasma. 
Also, it is known that plasma ejection leading to radio emission is known to occur at 
the onset of the low-hard spectral state of variable X-ray sources.

\noindent (iv) The spring 2003 event seems to be  an outlier in the $L_x - P$ diagram 
of Meyer-Hofmeister (2004) characterizing X-ray nova outbursts. Specifically,
given its relatively long orbital period of 1.5 days, it is somewhat surprising for a low peak
luminosity, $\sim$2\% $L_{edd}$,  low-hard-state outburst to occur.

\noindent (v) Given the complex outburst history of this source, as noted in item (i),
combined with the approximate FRED profile of the spring 2003 outburst, we 
have conjectured on the possibility that multiple outburst mechanisms may be
at play. In this context, we modeled the light curve profile as a discrete
accretion event, undergoing diffusive propagation through the disk. This leads
to a reasonable representation of the data, from which we find a $\sim70$-day viscous
timescale, and a relatively high mean viscosity.

\section {ACKNOWLEDGEMENTS}

This work made use of the NASA High-Energy Astrophysics
Research Archive Center (HEASARC), and the NASA {\it INTEGRAL}
Guest Observer Facility.  The Off-line Scientific Analysis (OSA) software available from the
{\it INTEGRAL} Science Data Centre was used to perform much of the analysis in this work.

This work is based in part on observations with {\it INTEGRAL}, an ESA project with instruments 
and science data centre funded by ESA member states (especially the PI countries: 
Denmark, France, Germany, Italy, Switzerland, Spain), Czech Republic and Poland, 
and with the participation of Russia and the USA.

\newpage

\begin{deluxetable}{cccccccccccccc}
\tablewidth{0pt}
\tabletypesize{\tiny}
\tablecaption{Best Fit Parameters for COMPST, COMPTT, and PEXRAV Models\tablenotemark{a}}
\tablehead{
\colhead{}  &
\multicolumn{3}{c}{COMPST\tablenotemark{b}} &
\colhead{}&
\multicolumn{4}{c}{COMPTT\tablenotemark{c}} &
\colhead{}&
\multicolumn{4}{c}{PEXRAV\tablenotemark{c,d}} \\
\colhead{Rev} &
\colhead{kT} &
\colhead{$\tau$} &
\colhead{$\chi^2$} &
\colhead{}&
\colhead{kT} &
\colhead{$\tau_{\rm p}$} &
\colhead{T$_0$}&
\colhead{$\chi^2$} &
\colhead{}&
\colhead{$\Gamma$} &
\colhead{${\rm E_{\rm cut}}$} &
\colhead{R}&
\colhead{$\chi^2$} \\
\colhead{} &
\colhead{(keV)} &
\colhead{} &
\colhead{} &
\colhead{} &
\colhead{(keV)} &
\colhead{} &
\colhead{(keV)} &
\colhead{} &
\colhead{} &
\colhead{} &
\colhead{(keV)} &
\colhead{} &
\colhead{}
}
\startdata
55 &  41.2$_{-1.7}^{+1.9}$ & 3.58$_{-0.13}^{+0.13}$ &  334.9 & & 52.6$_{-4.2}^{+6.2}$ &
 1.38$_{-0.15}^{+0.12}$ &  0.58$_{-0.14}^{+0.10}$ & 292.7 &  & 1.56$_{-0.05}^{+0.05}$ & 
 441.4$_{-122.0}^{+242.2}$ & 0.40$_{-0.22}^{+0.24}$& 303.7  \\ 
  \\
57 &  41.3$_{-1.7}^{+1.9}$ & 3.44$_{-0.14}^{+0.14}$ &  325.1 & & 49.4$_{-3.4}^{+4.5}$ &
 1.42$_{-0.12}^{+0.11}$ &  0.56$_{-0.16}^{+0.11}$ & 287.7 & & 1.53$_{-0.06}^{+0.06}$ & 
 310.3$_{-73.7}^{+130.9}$ & 0.31$_{-0.23}^{+0.26}$& 313.1 \\ 
  \\
60 &  36.4$_{-1.2}^{+1.3}$ & 3.92$_{-0.12}^{+0.13}$ &  389.4 & & 43.2$_{-2.4}^{+2.5}$ &
 1.64$_{-0.10}^{+0.11}$ &  0.34$_{-0.34}^{+0.39}$ & 295.0 & & 1.42$_{-0.05}^{+0.05}$ & 
 201.0$_{-31.5}^{+44.2}$ & 0.18$_{-0.18}^{+0.20}$& 300.9
\enddata
\tablenotetext{a}{$N_{\rm H} \equiv 9 \times 10^{21}$, errors are 90\% confidence}
\tablenotetext{b}{206 Degrees of freedom}
\tablenotetext{c}{205 Degrees of freedom}
\tablenotetext{d}{$i  \equiv 72.6^{\circ}$, see \citet{oros02}}
\end{deluxetable}

\begin{deluxetable}{cccc}
\tablewidth{0pt}
\tabletypesize{\tiny}
\tablecaption{Best Fit Model 3-300 keV Fluxes\tablenotemark{a,b}}
\tablehead{
\colhead{}  &
\colhead{COMPST} &
\colhead{COMPST} &
\colhead{COMPST} 
}
\startdata
55 &  7.68 &  8.07  &   8.32 \\ 
  \\
57 &  9.73 & 9.95 & 10.2\\ 
  \\
60 &  9.20 & 9.59 & 9.89
\enddata
\tablenotetext{a}{$10^{-9}\ {\rm ergs\ cm^{-2}\ s^{-1}}$}
\tablenotetext{b}{Unabsorbed, using ISGRI normalization}
\end{deluxetable}

\newpage
\begin{figure}
\plotone{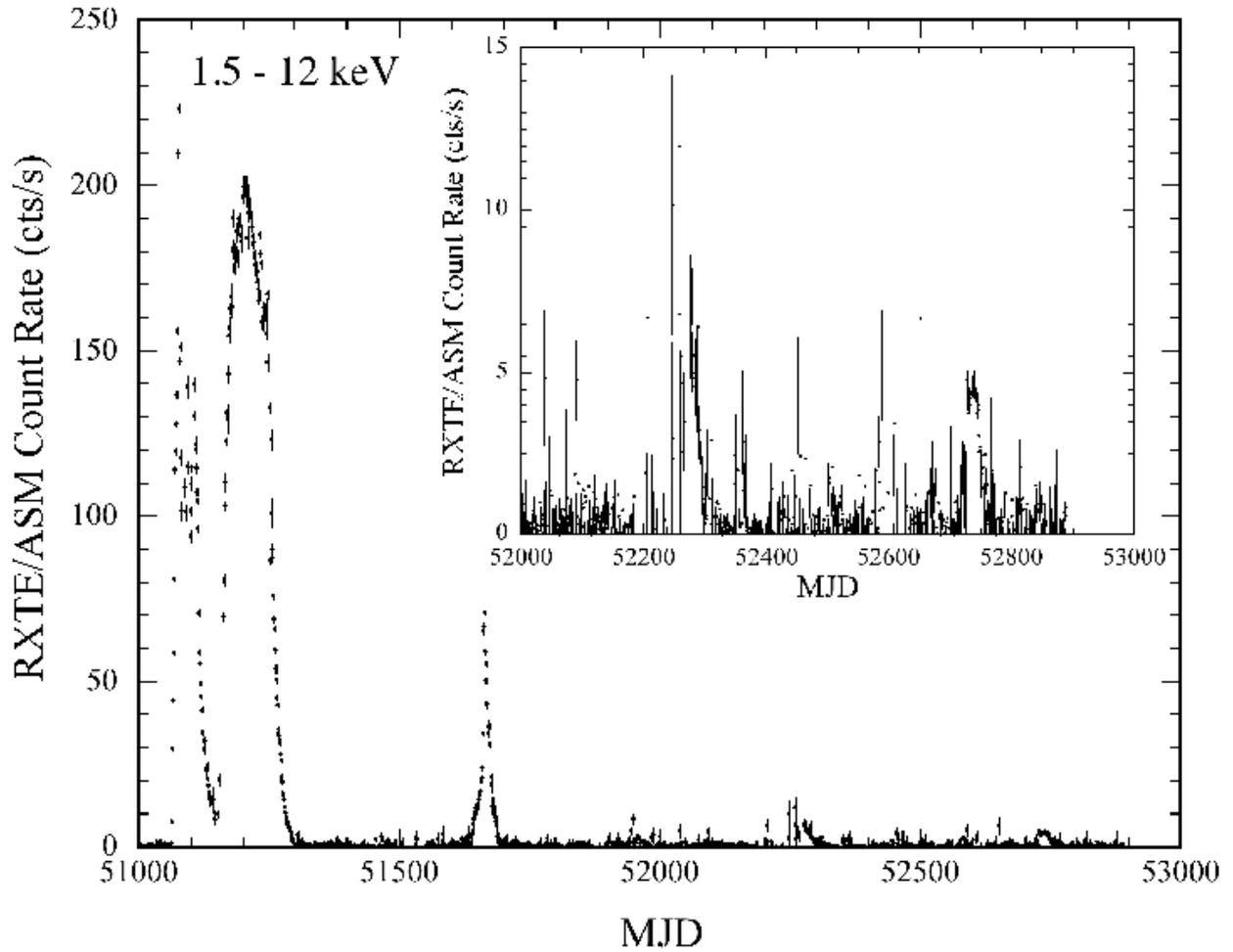}
\caption{Long term RXTE/ASM (1.5 - 12 keV) light curve. Vertical dotted
line overlays the March/April 2003 event.  The prominient outbursts of 1999
and 2000 have amplitudes which far exceed the minor flares 
which seem to occur with a fair degree of regularity after about MJD 52000.
  These small flares are more clearly  illustrated in the inset.}
\end{figure}

\newpage
\begin{figure}
\epsscale{0.4}
\plotone{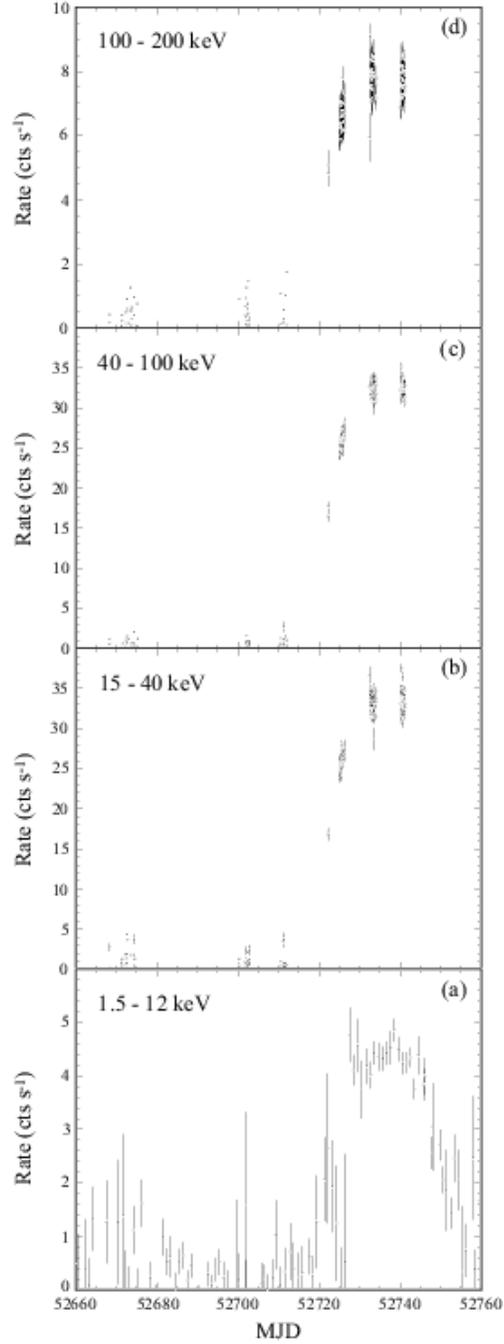}
\caption{Burst light curves for the (a) RXTE/ASM 1.5-12 keV, (b) IBIS/ISGR 15-40keV,
(c) IBIS/ISGRI 40-100 keV, and (d) IBIS/ISGRI 100-200 keV energy bands. The 3 INTEGRAL 
observations with the best statistics provide reasonable coverage of the late-rise 
phase and outburst peak. The lack of any extreme spectral evolution in the hard-X-ray
regime is evident from comparing the three INTEGRAL energy bands.
}
\end{figure}

\newpage
\begin{figure}
\epsscale{1.0}
\plotone{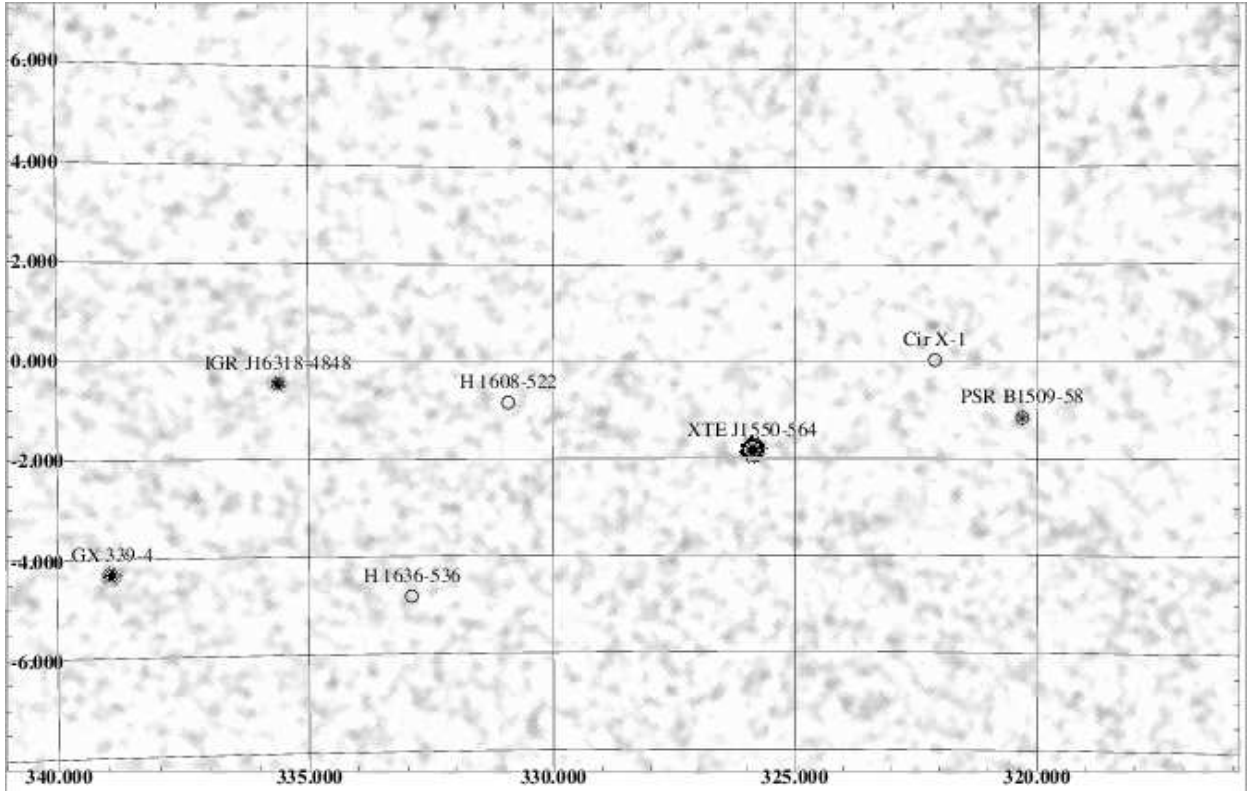}
\caption{40-100 keV IBIS/ISGRI significance image for INTEGRAL spacecraft 
revolution 57 (approximately MJD 52733).
Note that  IGR J16318-484 and GX 339-4 are also active 
during this time period. }
\end{figure}

\newpage
\begin{figure}
\epsscale{0.65}
\plotone{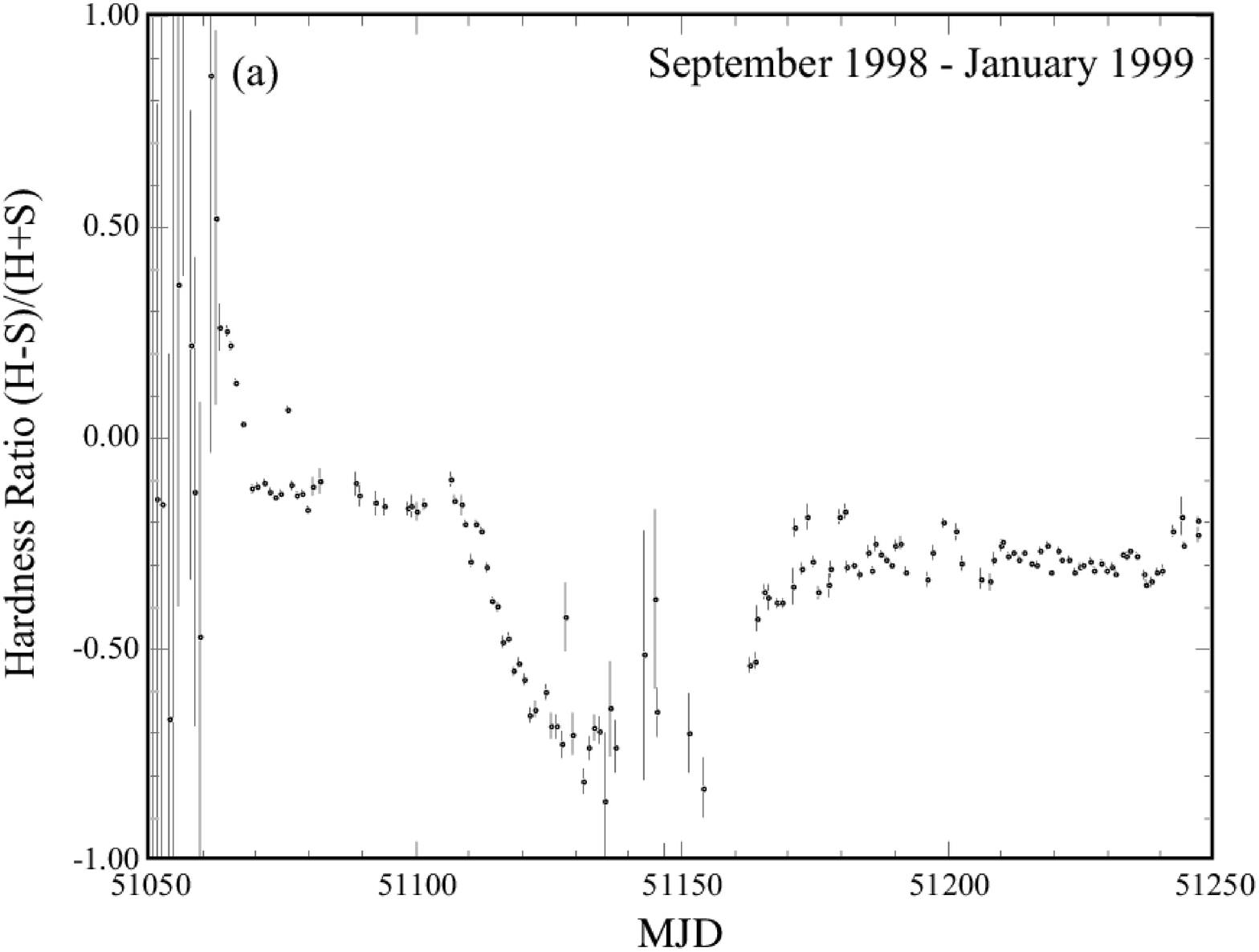}
\plotone{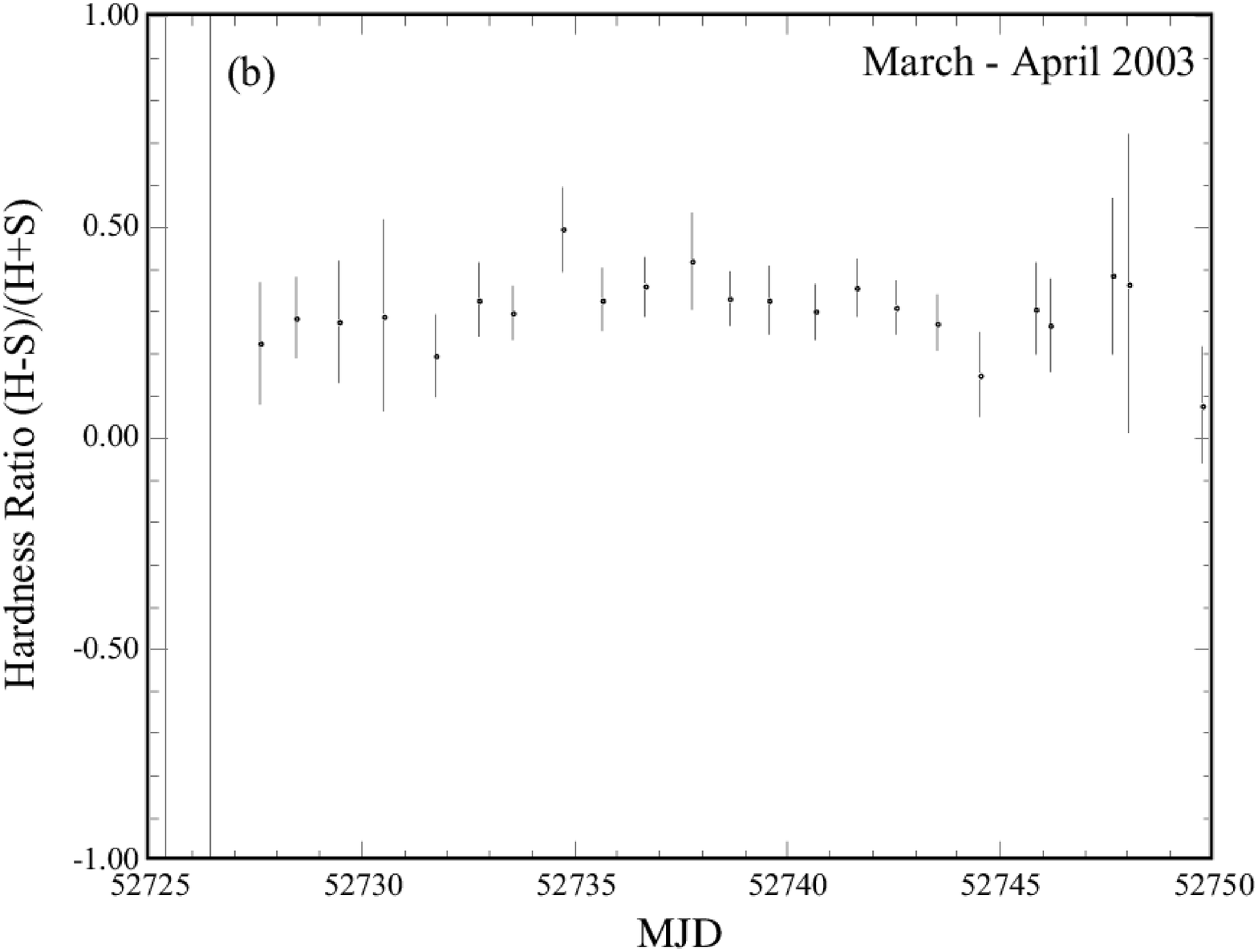}
\caption{A comparison of the 5-12 keV (H) to 1.5-3 keV (S) hardness ratios for the major January 1999 outburst and
the March-April 2003 outburst.  Here we have defined the hardness ratio as (H-S)/(H+S).  Note that the ratios are
roughly constant in both cases but at very different values.}
\end{figure}

\newpage
\begin{figure}
\epsscale{1.0}
\plotone{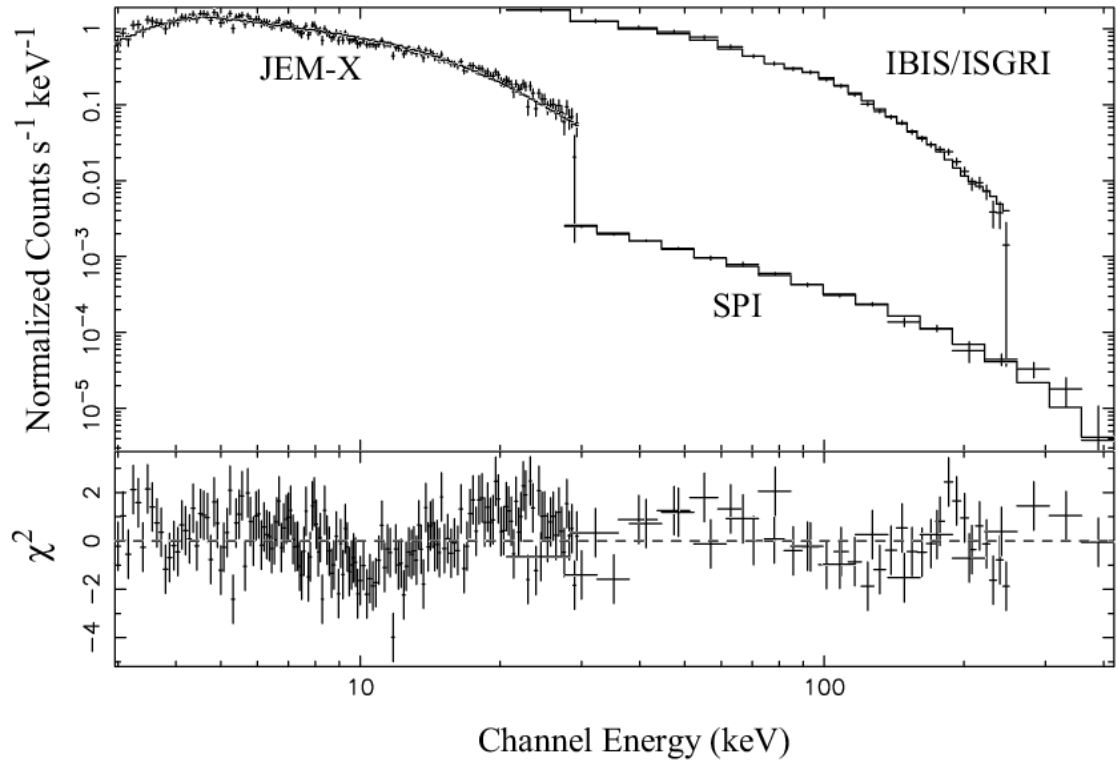}
\caption{The JEM-X, IBIS/ISGRI, and SPI data for XTE J1550-564 for revolution 57 as well as the best-fit
COMPTT model.  Also shown are the deviations of the model from the data in units of $\chi^2$.}
\end{figure}

\newpage
\begin{figure}
\epsscale{1.0}
\plotone{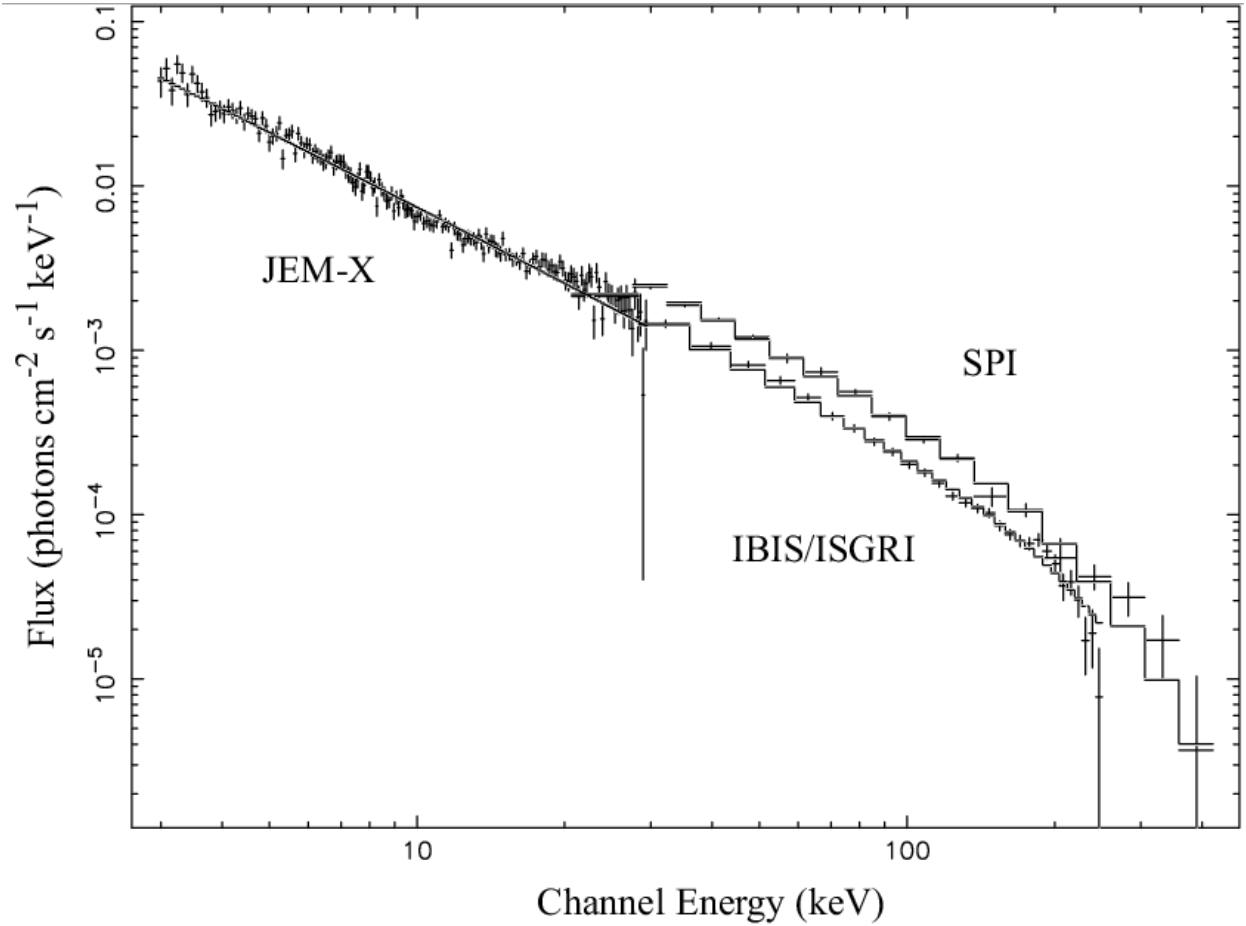}
\caption{The JEM-X, IBIS/ISGRI, and SPI unfolded data for XTE J1550-564 for revolution 57 as well as the best-fit
COMPTT model. The instrumental cross-calbration discrepencies are evident (see text for discussion). }
\end{figure}

\newpage
\begin{figure}
\epsscale{1.0}
\plotone{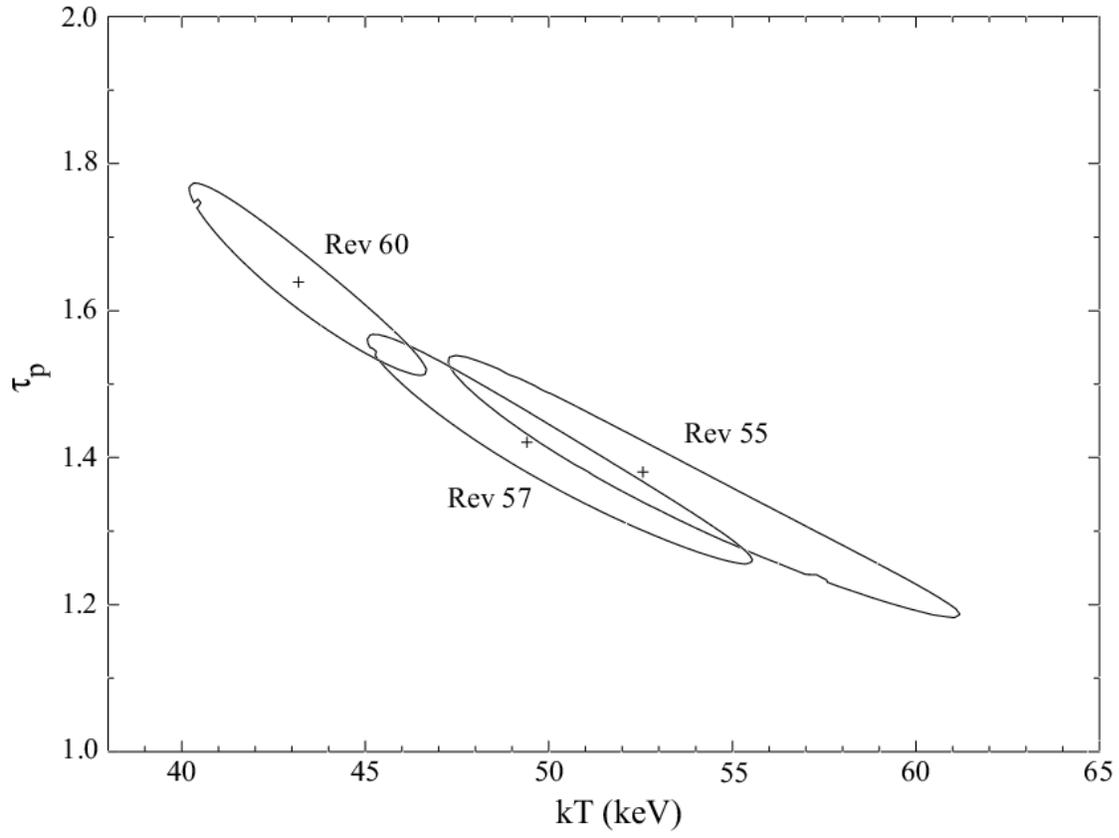}
\caption{Here we show the evolution of 90\% confidence error contours for the parameters kT and $\tau_{\rm p}$ of the
COMPTT model.   Note the evidence for spectral evolution to lower kT and higher $\tau_{\rm p}$ with time.}
\end{figure}

\newpage
\begin{figure}
\epsscale{1.0}
\plotone{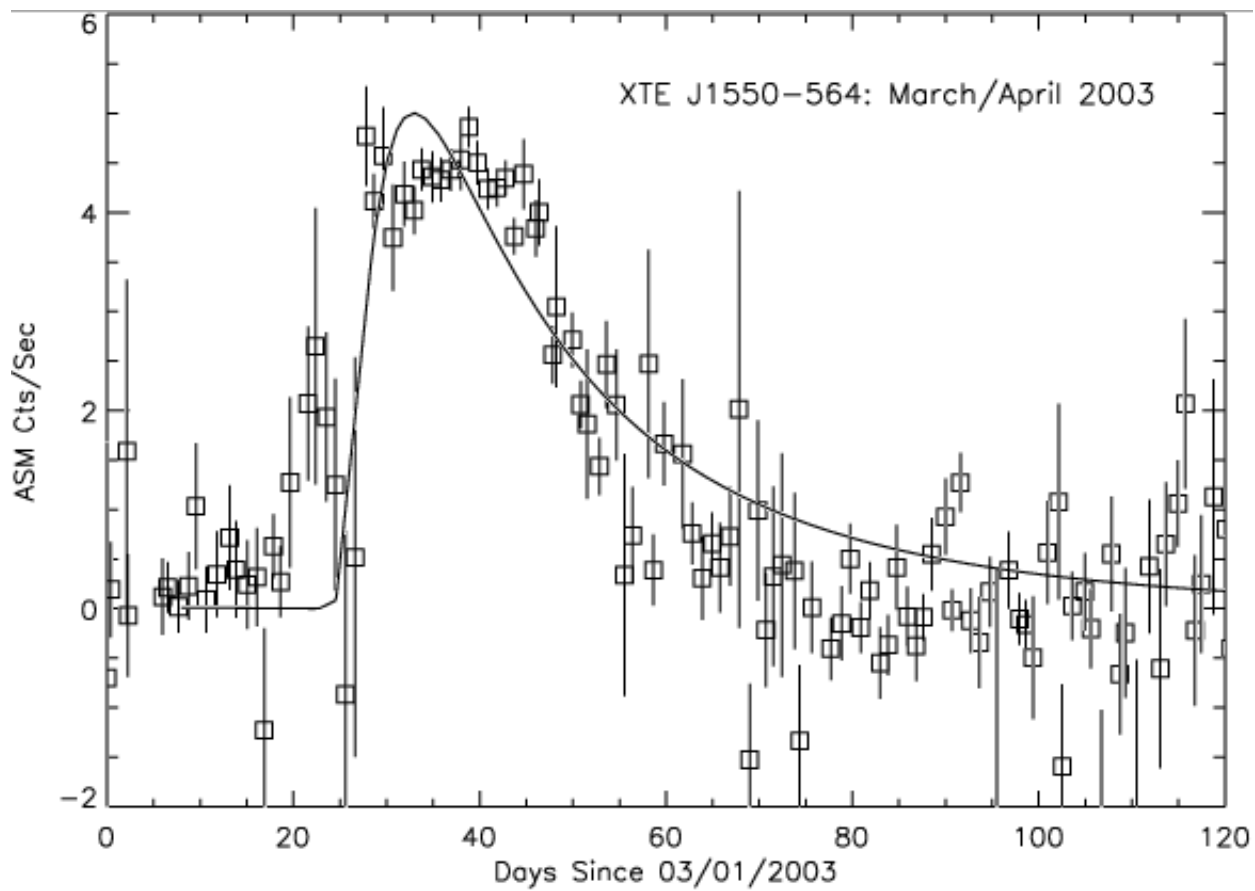}
\caption{The ASM summed-band (i.e. 3-12 keV) is plotted as detector count rate, versus days with a zero point
at March 1, 2003. For perspective, the ASM registers about 75 cts/s for the Crab.
The FRED-like profile, possibly even with a secondary maximum event (see Chen, Shrader
\& Livio 1997) about 20 days after the initial rise make it quite distinct from the major outbursts of 1999 
and 2000 in terms of both shape and amplitude.
The smooth curve overlaying
the data results to least-chisquare fit to the data based on the diffusive-disk propogation model of
Wood et al (2001). That model is predicated on a discrete accretion-rate increase 
 }
\end{figure}

\end{document}